# HFiTT – Higgs Factory in Tevatron Tunnel (Rev. 3)

Weiren Chou[1], Gerard Mourou[2], Nikolay Solyak[1], Toshiki Tajima[3], Mayda Velasco[4]

[1] Fermilab, USA
[2] École Polytechnique, France
[3] University of California at Irvine, USA
[4] Northwestern University, USA

May 20, 2013

White Paper for the 2013 US HEP Community Summer Study (Snowmass2013)

**Introduction**

Among various options for a Higgs factory [1], a photon collider has the distinct advantage of the lowest energy requirement for an electron beam. This advantage is especially important for a circular Higgs factory, in which the synchrotron radiation power increases to the fourth power of the electron energy. For an $e^+e^-$ collider, the minimum required energy per beam is 120 GeV, while for a photon collider it is 80 GeV. The corresponding ratio of synchrotron radiation power is 5 to 1. This makes it possible to consider building a photon collider at Fermilab, which will be named HFiTT, or *Higgs Factory in Tevatron Tunnel*. The layout is shown in Figure 1.

A photon collider is based on Inverse Compton Scattering (ICS) by shooting a low energy (~3.5 eV) laser beam into a high energy (10s of GeV) electron beam to generate a back-scattered high energy (10s of GeV) photon beam for collision. The cross section for $\gamma\gamma \to H$ is large and comparable to $e^+e^- \to ZH$ (~200 fb). Since this is an s-channel resonance, the required photon energy is low (63 GeV), corresponding to 80 GeV for an electron beam. Our design goal is 10,000 Higgs per year.

If we look to the far future of electron collider technology, very high gradient acceleration techniques such as plasma wakefield acceleration are asymmetric between electrons and positrons. It is therefore important to develop a technology that will allow physicists to accelerate electrons only and still access annihilation reactions with precisely understood point-like interactions. The photon collider fills this need. This project will not only carry out important measurements of the Higgs boson properties but will also demonstrate the unique technologies needed to construct photon collider experiments at higher energies.

**Physics**

There have been several publications on the physics of a photon collider. [2-4] The strength of a Higgs factory based on a photon collider is its ability to study CP admixture and violation of the Higgs to better than 1%. This cannot be done by any other machine. Such a study can be carried out by using either circularly or linearly polarized photon beams. The unique attribute of a photon collider is the initial coupling to a pair of photons. Higgs can be observed in, e.g. the $\bar{b}b$ final state, following which, using the $\bar{b}b$ partial width measured at another machine, the $H \to \gamma\gamma$ partial width can be extracted to a precision of 1%. This quantity is of particular interest because this decay proceeds through an inclusive loop that can potentially reveal heavier charged particles into which the Higgs cannot decay directly. A photon collider also has the ability of measuring precisely the rates for several additional processes, such as $\gamma\gamma \to H \to WW^*, ZZ^*$ and perhaps $\tau^+\tau^-$ and $gg$. It can also study non-standard Higgs decays. These contributions – high precision measurement of couplings to SM particles and measurement of non-standard Higgs decays – go beyond the LHC. The precisions on Higgs coupling to various particles are listed in Table 1.

In addition to the Higgs physics, HFiTT can also study $e^-e^-$ collisions at 80 GeV. This experiment can make precision measurement of the energy dependence of $\sin^2\theta_W$ to understand the inconsistency between the SLD and LEP measurements at the Z-pole. This is possible from measurements of the Left-Right asymmetry in polarized Møller scattering. The measurements will be in the 10 to 115 GeV range.

**Accelerator**

Our design is similar to a recirculating electron linac. As illustrated in Figure 1, there are eight RF stations. Each station consists of five ILC-type cryomodules. Each module has nine ILC-type SRF cavities. Each station occupies a straight section of 70 meters and provides 1.25 GV acceleration voltage. The total RF is 10 GV. The required RF power is 27 MW (24 MW for beams and 2.3 MW for synchrotron radiation), which leads to 75 kW per coupler. The dynamic heat load at 2°K is ~850 W per cryomodule for CW operation. The wall power for cryogenics is about 25 MW.

There are eight recirculating beam lines for each electron beam. These beam lines are made of Recycler-type permanent magnets. Each section has different field strength, which is determined by the beam energy in that section. The maximum bending field is 3.3 kG. If a scanning of a few GeV around the peak collision energy is desired, the magnets in the final arcs can use powered ones to provide variable field strength.

The Tevatron tunnel has a radius of 1,000 m and a circumference of 6,283 m. The tunnel size is 3.048 m (10 feet) wide and 2.438 m (8 feet) high. The Tevatron used 53.2 m long straight sections. But these straight sections can be readily extended to 70 m or longer.

Two electron beams will be injected in opposite directions. Each beam will circulate eight times and be accelerated to 80 GeV before they are directed to the experimental hall. This hall has a large detector and also a fiber laser system. Two laser beams will collide with the two electron beams at the conversion point (CP) respectively and generate two high energy (~64 GeV) photon beams for head-on collision at the interaction point (IP) in the center of the detector. The spent electrons need to be dumped before the IP. The dumping scheme is a major R&D for a photon collider.

The main accelerator and beam parameters are listed in Table 2. Many of the electron beam parameters are similar to those of an ILC-based photon collider. But the collision frequency is 47.7 kHz, which is about a factor 3 higher than an ILC-based photon collider. The electron beam intensity is $2 \times 10^{10}$ per bunch. For each circulating beam, the current is 0.15 mA for each line or 1.22 mA for 8 lines.

The injection requires a low emittance, highly polarized (80%) electron gun for flat beams. The polarization must be controlled carefully to go around a bend and end up in the right state in the ring. We can also envision that the injection beam comes from ASTA or Project X if their beams are available for HFiTT.

**Laser**

A major challenge for HFiTT is the required laser beam, which must have high repetition rate (47.7 kHz) and high average power (240 kW). A recent breakthrough in fiber laser technology demonstrates that such a laser system is indeed within reach. Reference [5] shows that thousands of fiber channels can be combined coherently to produce a pulse with an energy of >10 J at a repetition rate of ~10 kHz (Figure 2). Such a system would meet our needs. There are, of course, some critical issues that have to be addressed before one can claim the required laser technology is in hand, for instance, protection of the optical components near the CP.

**Detector**

After the completion of the Tevatron experiment, the DZero detector is in "frozen" state. [6] A number of its sub-systems could be resurrected and reused. A new tracker and precision vertex system will be needed.

**Cost Consideration**

This proposal is at an early stage and it is premature to discuss about its total cost. However, it will be useful to provide cost references for major systems based on the ILC study and Recycler experience.

- 40 cryomodules. Cost – $2-3 million each according to the ILC cost estimate. (As a comparison, the ILC would need ~1,700 cryomodules.)
- 27 MW of RF power. Assuming 50% efficiency, one needs 54 MW of wall power for RF. Cost – $5 million per MW according to the ILC cost estimate.
- 25 MW of wall power for cryogenics. Cost – about 2/3 of the ILC cryogenics.
- 16 permanent magnet beam lines. Cost reference – the Recycler permanent magnet total cost was $3.2 million.
- 2×240 kW laser system. Assuming wall plug efficiency of 30%, compressor efficiency of 50%, diode price of €5/W and the rule of thumb that "3 times the diode cost equals the cost of the full system," the laser system will cost ~€50M, or $65 million.
- Civil – the Tevatron tunnel, CDF and DZero experimental halls, service buildings and utilities can be reused to minimize the civil cost.

**Acknowledgment**

We want to thank Bogdan Dobrescu and Michael Peskin for reading this document carefully and making important and valuable comments on physics and other sections. We thank Tom Peterson for the cryogenic calculation. We thank Chris Barty for stimulating discussions on laser architecture. We thank Wilhelm Becker, Howard Bryant, Enrico Brunetti and Dino Jaroszynski for the ICS discussions and calculations. We also thank

**Table 1:** Precision of measurements to be performed at HFiTT after 5 years of data taking

| Measurement | Precision after 5 years of operation | Comment |
|---|---|---|
| $\Gamma_{\gamma\gamma} \times \text{Br}(h \to \bar{b}b)$ | 0.01 | |
| $\Gamma_{\gamma\gamma} \times \text{Br}(h \to WW^*)$ | 0.03 | Leptonic decays only |
| $\Gamma_{\gamma\gamma} \times \text{Br}(h \to \gamma\gamma)$ | 0.12 | |
| $\Gamma_{\gamma\gamma} \times \text{Br}(h \to ZZ^*)$ | 0.06 | One Leptonic and one hadronic decay |
| $\Gamma_{\gamma\gamma} \times \text{Br}(h \to Z\gamma)$ | 0.20 | Leptonic and hadronic decays for $Z$ |
| $\Gamma_{\gamma\gamma} \times \text{Br}(h \to \tau^+\tau^-)$ | - | Work in progress |
| $\Gamma_{\gamma\gamma} \times \text{Br}(h \to \bar{c}c)$ | - | Work in progress |
| $\Gamma_{\gamma\gamma} \times \text{Br}(h \to gg)$ | - | Work in progress |
| $\Gamma_{\gamma\gamma} \times \text{Br}(h \to \mu^+\mu^-)$ | 0.38 | |
| $\Gamma_{\gamma\gamma}$ | 0.02 | Using Br($h \to \bar{b}b$) as input |
| $\Gamma_{\text{total}}$ | 0.13 | Using Br($h \to \bar{b}b$) as input |
| $H_{tt}$ Yukawa coupling | 0.04 | Indirect from $\Gamma_{\gamma\gamma}$ |
| Mass measurement | 60 MeV | From $h \to \gamma\gamma$ |
| CP Asymmetry using $h \to \bar{b}b$ | <0.01 | |
| CP Asymmetry using $h \to WW^*$ | 0.04 | |

**Table 2:** HFiTT accelerator and beam parameters.

| Top level parameters | | |
|---|---|---|
| Collision energy (center of mass) | GeV | 126 |
| Luminosity (per IP) | $10^{34}$ cm$^{-2}$ s$^{-1}$ | 0.5 |
| definition of luminosity | | $\gamma\gamma$ > 125 GeV |
| Luminosity for e-e- | $10^{34}$ cm$^{-2}$ s$^{-1}$ | 3.2 |
| No. of IP | | 1 |
| No. of Higgs per year (per IP) | | 10,000 |
| Circumference | km | 6.28 |
| P(wall) | MW | 80 |
| Polarization e- | | 80% |
| Polarization $\gamma$ | | 90% (lum. peak) |
| **Accelerator parameters** | | |
| Machine radius | m | 1000 |
| Revolution frequency | kHz | 47.7 |
| Bending radius | m | 800 |
| Bending field | kG | 0.05 –3.3 |
| RF voltage | GV | 10 |
| RF power (total) | MW | 26.7 |
| RF power (per coupler) | kW | 75 |
| No. of recirculating arcs | | 8 |
| **Electron beam parameters** | | |
| Beam energy | GeV | 80 |
| Energy loss per turn (at 80 GeV) | GeV | 4.53 |
| Number of electrons per bunch | $10^{10}$ | 2 |
| Number of bunches | | 1 |
| Collision frequency | kHz | 47.7 |
| Circulating beam current | mA | $1.22 \times 2$ |
| Collision beam current | mA | $0.15 \times 2$ |
| Beam power | MW | $12.2 \times 2$ |
| Synchrotron radiation power | MW | 2.3 |
| $\varepsilon_{x,n}$ | mm-mrad | 10 |
| $\varepsilon_{y,n}$ | mm-mrad | 0.03 |
| beta_x CP | mm | 4.5 |
| beta_y CP | mm | 5.3 |
| $\sigma_x$, CP | nm | 535 |
| $\sigma_y$, CP | nm | 32 |
| $\sigma_z$, CP | mm | 0.35 |
| sigma_E IP | % | 0.22 |
| **Laser beam parameters** | | |
| Wavelength | μm | 0.351 |
| Pulse energy | J | 5 |
| Repetition rate | kHz | 47.7 |
| Peak power | TW | 1.5 |
| Average power | kW | 240 |
| Rayleigh length | mm | 0.63 |
| $\sigma_x$, CP | nm | 4200 |
| $\sigma_y$, CP | nm | 4200 |
| $\sigma_z$, CP | mm | 0.45 |
| IP<->CP distance | mm | 1.4 |
| Laser-beam crossing angle | mrad | |
| **$\gamma$ beam parameters** | | |
| n_gamma | $10^{10}$ | 1 (primary) |
| $\sigma_x$, IP | nm | 480 |
| $\sigma_y$, IP | nm | 10 |

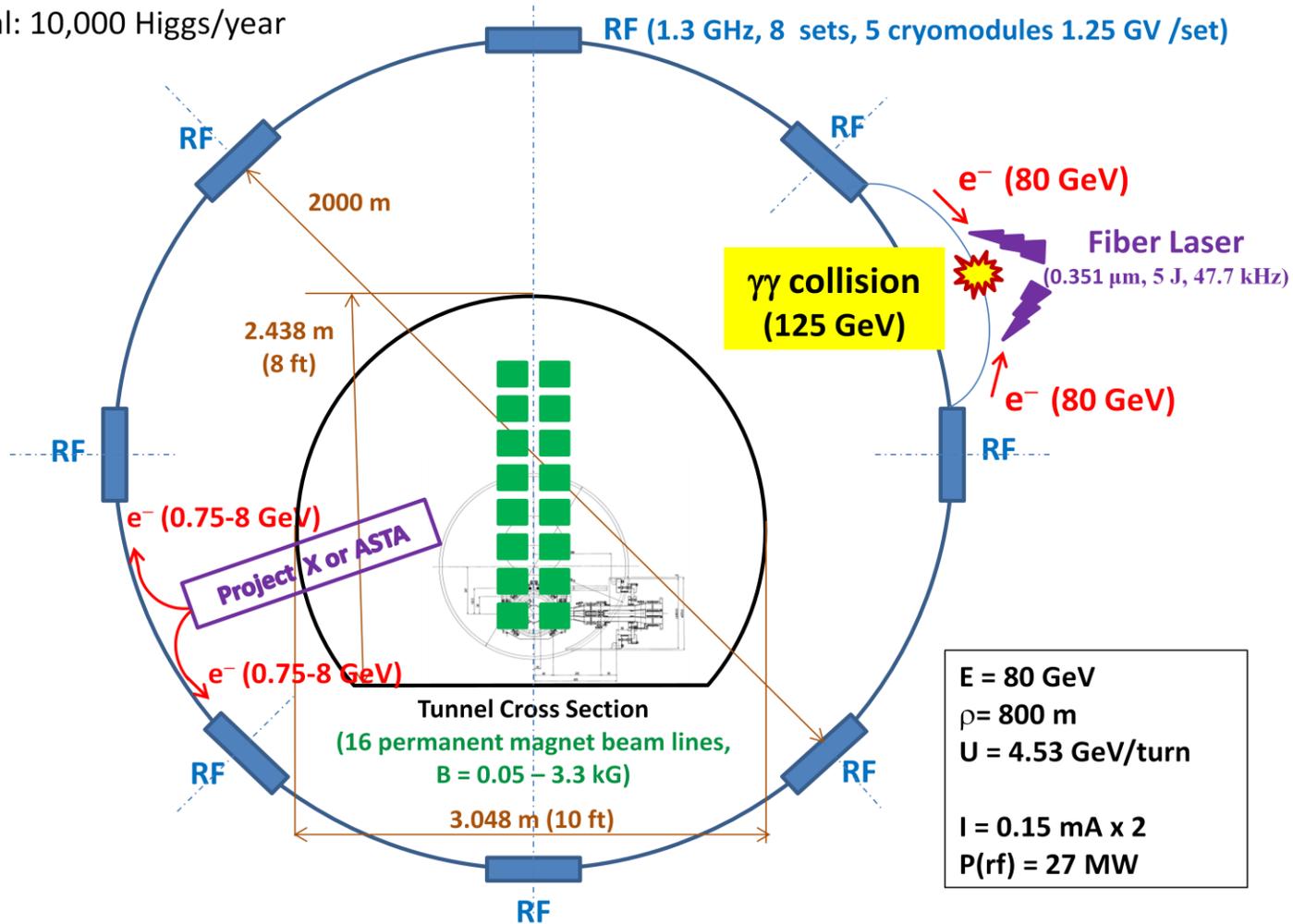

**Figure 1:** Layout of HFiTT.

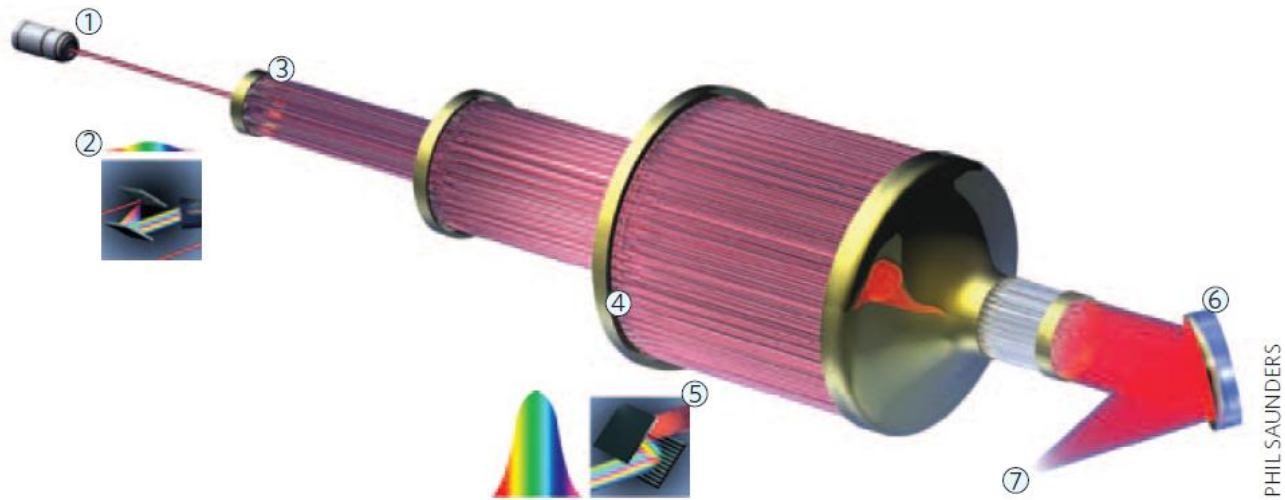

**Figure 2:** Principle of a coherent amplifier network (CAN) based on fiber laser technology. An initial pulse from a seed laser (1) is stretched (2), and split into many fibre channels (3). Each channel is amplified in several stages, with the final stages producing pulses of ~1 mJ at a high repetition rate (4). All the channels are combined coherently, compressed (5) and focused (6) to produce a pulse with an energy of >10 J at a repetition rate of 10 kHz (7). [5]